\documentclass[]{spie}  

\usepackage{amsmath,amsfonts,amssymb}
\usepackage[T1]{fontenc} 
\usepackage[]{graphicx} 
\usepackage{multirow}  
\usepackage[colorlinks=true, allcolors=blue]{hyperref}

\def\CN2{\mbox{$C_N^2 $}}

\def\CT2{\mbox{$C_T^2 $}}

\def\sigmal2{\mbox{$\sigma ^{2}_{I} $}}

\title{Towards an automatic system for monitoring of \CN2 and wind speed profiles with GeMS}

\author[a]{Elena Masciadri}
\author[b]{Benoit Neichel}
\author[c]{Andres Guesalaga}
\author[a]{Alessio Turchi}
\affil[a]{INAF - Osservatorio Astrofisico di Arcetri, L.go E. Fermi 5, 50125  Florence, Italy}
\affil[b]{Aix Marseille Universit\'e, CNRS, LAM (Laboratoire d'Astrophysique de Marseille) UMR 7326, 13388, Marseille, France}
\affil[c]{Pontificia Universidad Catolica de Chile, Santiago, Casilla 7820436, Chile}

\authorinfo{Send correspondence to Elena Masciadri - e-mail: masciadri@arcetri.astro.it}

\begin{document} 
\maketitle

\begin{abstract}
Wide Field Adaptive Optics (WFAO) systems represent the more sophisticated AO systems available today at large telescopes. One critical aspect for these WFAO systems in order to deliver an optimised performance is the knowledge of the vertical spatiotemporal distribution of the \CN2 and the wind speed. 
Previous studies (Cortes et al., 2012[\cite{cortes2012}]) already proved the ability of GeMS (the Gemini Multi-Conjugated AO system) in retrieving \CN2 and wind vertical stratification using the telemetry data. To assess the reliability of the GeMS wind speed estimates a preliminary study (Neichel et al., 2014[\cite{neichel2014a}]) compared wind speed retrieved from GeMS with that obtained with the atmospherical model Meso-Nh on a small sample of nights providing promising results. The latter technique is very reliable for the wind speed vertical stratification. The model outputs gave, indeed, an excellent agreement with a large sample of radiosoundings ($\sim$ 50) both in statistical terms and on individual flights (Masciadri et al., 2013[\cite{masciadri2013}]).
Such a tool can therefore be used as a valuable reference in this exercise of cross calibrating GeMS on-sky wind estimates with model predictions.
The main results of Neichel et al. (2014) analysis showed that, on a great number of cases, GeMS could reconstruct very good wind speed estimates. At the same time it has been put in evidence, on a number of cases, not negligible discrepancies from the atmospherical model. However we observed that these discrepancies strongly decreased or even disappear if GeMS data reduction is done with the {\it a priori} knowledge of the wind speed stratification provided by the model Meso-Nh. Basically the {\it a priori} knowledge helped the data reduction of GeMS acquisitions. In this contribution we achieved a two-fold results: (1) we extended analysis on a much richer statistical sample ($\sim$ 43 nights), we confirmed the preliminary results and we found an even better correlation between GeMS observations and the atmospherical model with basically no cases of not-negligible uncertainties;  (2) we evaluate the possibility to use, as an input for GeMS, the Meso-Nh estimates of the wind speed stratification in an operational configuration. Under this configuration these estimates can be provided many hours in advanced with respect to the observations and with a very high temporal frequency (order of 2 minutes or less). Such a system would have a set of advantages:  (a) to implement inside GeMS a total temporal and spatial coverage of the wind speed over $\sim$ 20 km and all along the night not only in real-time but in advance of a few hours, (b) to improve the detection of the \CN2 vertical stratification from GeMS because a good wind speed estimation would improve the quality of the cross-correlation peaks detection, (c) the possibility to bypass the complex (and not necessarily reliable) procedures necessary to automatise the wind speed estimate of GeMS due to the relatively low vertical resolution of the system. 
Such a study can obviously be considered as a demonstrator for multiple operational AO and WFAO systems (AOF, LINC-NIRVANA, RAVEN, ...) of present top-class telescopes and for the forthcoming generation. It might have, therefore, an interest for the AO community well beyond the improvement of GeMS performance.
\end{abstract}


\section{INTRODUCTION}
\label{sec:intro}  

Wide Field Adaptive Optics (WFAO) systems represent the more sophisticated AO systems available today at large telescopes. They include different typologies of adaptive optics: the ground layer adaptive optics (GLAO), the multi-conjugated adaptive optics (MCAO), the laser tomography adaptive optics (LTAO) and the multi-object adaptive optics (MOAO). The common feature for all WFAO systems is that they significantly increase the field of view (FoV) of the AO-corrected images and the fraction of the sky that can benefit from such correction. WFAO have been conceived, indeed, to correct the perturbed wavefront on field of views (FoV) of the order of a few arc-minutes instead of the typical 10 arcsec of the single conjugated adaptive optics system (SCAO). In WFAO systems light from several GSs is used to probe the instantaneous three-dimensional phase perturbations, and the turbulence volume is then reconstructed by solving an inverse problem. This technique, called atmospheric tomography, was first introduced by Tallon \& Foy[\cite{tallon1990}] and later improved by Johnston \&  Welsh[\cite{johnston1994}], Ellerbroek[\cite{ellerbroek1994}], and Fusco et al.[\cite{fusco2001}], among others. 

Performances of all WFAO systems strongly depends on the knowledge of the 3D turbulence distribution. Erroneous estimates of the atmospheric conditions deeply deteriorates the tomographic performances [\cite{neichel2008}]. Some fundamental limitations of the specific system (the unseen modes and the unseen turbulence) depend on the geometry of the system. An error on the height of the reconstructed layers affects the regularisation of modes. The height of the reconstructed layers provided to the AO system determines the sensitivity of the reconstructors to unseen frequencies. If the geometry does not fit with the real one, the applied regularisation will act in inefficient way. The impact of the model error has been studied in different papers: Conan et al.[\cite{conan2001}] and Tokovinin \& Viard[\cite{tokovinin2001}] studied how the reconstruction error depends on an error on the \CN2 profile; LeLouarn and Tallon[\cite{miska2002}] explored how the reconstruction of just a set of turbulent layers at specific altitudes was related to the whole turbulence volume; Fusco et al.[\cite{fusco1999}] studied the effect of a reconstruction on a limited number of equivalent layers (ELs); Costille \& Fusco[\cite{costille2012}] studied the impact of sampling the input turbulence for an application to ELTs. All these studies indicate that the larger the FoV, the more sensitive is the AO system to errors in the identification of the height of the layers. Besides the \CN2, the wind speed and direction are also particularly critical for the observations supported by WFAO systems. Temporal errors in adaptive optics are due to turbulence evolution during the time delay between the time in which the perturbed wavefront is measured and the time in which is corrected, the so called control cycle. It has been proved that the prediction of the turbulence state on this time scale (order of 1 KHz) can improve in not negligible way the performances of the wavefront correction. Under a frozen turbulence conditions the prediction of the successive state of turbulent layer mainly depends on the knowledge of the wind speed and direction of the turbulent layer at the present time t[\cite{johnson2008}]. The authors proved that the percentage gain in Strehl ratio using predictive controller under these conditions can achieve 40\% in high wind and high noise conditions. The knowledge of the wind speed and direction of the different turbulent layers is therefore extremely critical to recover the phase state in the delay time of the control cycle[\cite{gavel2002,poyneer2007,ammons2012}]. 

In this contribution we performs the comparison between wind speed measured by GeMS and reconstructed by the atmospherical model at mesoscale Meso-Nh on a rich statistical sample of 43 nights with the goal to provide a more complete analysis with respect to Neichel et al. 2014[\cite{neichel2014a}] and we present preliminary results related to the wind direction. In a previous study, indeed, Masciadri et al., 2013, 2015 [\cite{masciadri2013,masciadri2015}] showed that the non-hydrostatical mesoscale atmospherical model Meso-Nh was able to provide reliable wind speed and direction profiles on the whole troposphere and stratosphere (up to 20-25 km) above top-level astronomical sites. The statistical estimates done comparing model prediction with 50 radio-soundings distributed in different periods of the year indicates, for the wind speed, in the range 5-15~km a bias within 1~ms$^{-1}$ and a RMSE within 3 ms$^{-1}$. Above 15~km and in the 3-5~km range, the bias is within 2 ms$^{-1}$. A detailed analysis done on all individual 50 cases (Masciadri et al. 2013 - Fig.B1) tells us that an equivalent good performances are achieved by the model not only in statistical terms but also on each individual comparison model/radiosoundings. This tells us that this model can be used as a reference to check/validate the performances of GeMS estimates.

The purpose of this paper is to validate the reliability of the GeMS outputs and, at the same time, to evaluate the possibility to use the Meso-Nh ouputs as an input for GeMS. The procedure to automatise the GeMS measurement is indeed complex and it seems more advantageous to use the model estimate of the wind that is also more complete in space (the model covers the whole 20~km a.g.l. with 62 estimates instead of 2 or 3 points of GeMS) and time (a temporal frequency of two minutes). GeMS is, indeed, a WFAO with three DMs (0, 4.5 and 9km). GeMS is therefore much more efficient than a GLAO in obtaining high Strehl Ratio on a wide field of view correcting the turbulence on the whole atmosphere. But even so, it has been observed that the individual three couples WFSs-DMs can negatively be affected by some sudden increase of the wind speed or change of the wind direction and affect negatively the observations. We therefore can take advantage of the possibility to measure the wind speed with GeMS or, even better, to know in advance the wind speed with the Meso-Nh model outputs.
Knowing the wind direction and intensity in advance could be used for instance to update the control matrices, reducing or increasing the relative gains associated with each altitude layers. In case of predictive control, for instance for tip-tilt control, knowing the wind direction/intensity is a must.

\section{WIND SPEED FROM GeMS}
\label{sec:wind_speed_gems}

GeMS, the Gemini MCAO System, is the first multi-sodium based LGS AO system currently running on sky conceived for astronomic observation (Rigaut et al., 2014[\cite{rigaut2014}], Neichel et al., 2014[\cite{neichel2014b}]). It uses five LGSs distributed on a 1 arcmin constellation to measure and compensate for atmospheric distortions and delivers a uniform, close to diffraction-limited Near-Infrared (NIR) image over an extended FoV of 2arcmin. GeMS is a facility instrument for the Gemini South (Chile) telescope, therefore open to the community. It has been designed to feed two science instruments: GSAOI[\cite{mcgregor2004}], a 4k x 4k NIR imager and Flamingo2[\cite{elston2003}], a NIR multi-object spectrograph. 

\CN2 and wind speed profiles are retrieved using a method described in Neichel et al., 2014[\cite{neichel2014a}] using the SLODAR technique conceived for binaries (Wilson, 2002[\cite{wilson2002}]) extended to multiple laser guide stars WFAO systems running with Shack-Hartmann wave-front sensors[\cite{cortes2012,gilles2010,osborn2012}]. Turbulence strength is sampled with a vertical resolution $\Delta$h = d/$\theta$ where d is the sub-aperture projected on the pupil and $\theta$ the angular separation of different couple of stars selected among the guide stars of the asterisms taken in consideration. Different $\Delta$h can therefore be obtained depending on the selected binaries. Moreover, the cone effect typical of laser guide stars does that the the bins in altitude are not equally spaced (Eq.1[\cite{cortes2012}]). As an extension of the SLODAR method, the wind profiling consists in performing time-delayed cross-correlations between all possible combinations of slope measurements from the available sub-apertures of the WFS[\cite{wang2008}]. This profiling method provides information not only on the turbulence distribution in altitude but also on the wind velocity and direction for each layer and their dynamic evolution. The method has also been modified to include multiple LGS-WFSs and the handling of the cone and fratricide effects in the GeMS configuration[\cite{guesalaga2014}]. Uncertainties for the wind speed intensity (i.e. along x-axis) are estimated assuming $\pm$0.2 to 0.5 sub-aperture error, depending on the signal to noise ratio of the peak, and assuming $\pm$ 1 to 2 frames error depending on the translation speed of the peak. 

The method just described has been applied manually in this study i.e. the selection of the data from GeMS WFS has been done so that cross-correlation peaks (associated to turbulent layers) might be clearly and unambiguously detected. To identify the wind speed of a layer it is necessary to track its shift in the cross-correlation map. Tracking cross-correlation peaks are more easily identified when they have different velocity vectors and are not overlapped along the tracks. 

\section{WIND SPEED PROFILES FROM THE ATMOSPHERICAL MESO-SCALE MODEL MESO-NH}
\label{sec:wind_model}

Numerical calculations of wind speed and direction related to 43 nights have been obtained with a non-hydrostatical atmospherical mesoscale model called Meso-Nh[\cite{lafore1998}]. Meso-Nh can reconstruct the temporal evolution of three-dimensional meteorological parameters over a selected finite area of the globe. The system of hydrodynamic equations is based upon an anelastic formulation allowing for an effective filtering of acoustic waves.
It uses the Gal-Chen and Sommerville[\cite{gal75}] coordinates system on the vertical and the C-grid in the formulation of
[\cite{arakawa76}] for the spatial digitalisation.
It employs an explicit three-time-level leap-frog temporal scheme with a time filter[\cite{asselin72}].
It employs a one-dimensional 1.5 turbulence closure scheme[\cite{cuxart00}].
For this study we use a 1D mixing length proposed by Bougeault and Lacarr\`ere[\cite{bougeault89}].
The surface exchanges are computed using ISBA (Interaction Soil Biosphere Atmosphere) module[\cite{noilhan89}]. The model has been developed by the Centre National des Recherches M\'et\'eorologiques (CNRM) and Laboratoire d'A\'ereologie (LA) de l'Universit\'e Paul Sabatier (Toulouse). 
The package of the optical turbulence (\CN2) and integrated derived parameters (called Astro-Meso-Nh) have been developed by Masciadri et al.[\cite{masciadri1999a}]. Since then the Astro-Meso-Nh code has been improved and several studies mainly addressing the reliability of this technique in astronomical applications[\cite{masciadri2004,hagelin2011}] or in application to climatologic studies[\cite{masciadri2006}] have been carried out. A few among these studies have been directly focused on the ability of the Meso-Nh model in reconstructing the wind speed[\cite{masciadri2003,hagelin2010,masciadri2013,lascaux2011}]. The most recent version of the Astro-Meso-Nh used for this study is that described in[\cite{masciadri2013}].
Both the Meso-Nh code as well as the Astro-Meso-NH code for the optical turbulence are parallelised with OPEN-MPI.
The model can therefore be run on local workstations as well as on the High Performance Computing Facilities (HPCF) cluster of the 
European Centre for Medium Range Weather Forecasts (ECMWF), in parallel mode so as to gain in computing time.

We used the same model configuration used in Neichel et al.[\cite{neichel2014a}] with three domains in grid-nesting configuration having resolution of 10~km, 2.5km and 0.5~km. We simulated 43 nights for which GeMS telemetry data were available. Each simulation started at 18 UT of the day before and it was forced every 6 h with the analyses from the European Centre for Medium-Range Weather Forecasts (ECMWF). Simulations finished at 11 UT (07:00 LT) of the simulated day (for a total duration of 17 h). The statistics is computed only during night local time, from 00 UT to 09 UT. The vertical profiles extracted from the model computations are available each 2 minutes (temporal frequency).

\section{RESULTS}
\label{sec:results}

GeMS data reduction and simulation with the Meso-Nh model on the 43 nights have been performed in a completely independent way. As previously observed, the correlation between GeMS and Meso-Nh is substantially very good for the wind speed. Moreover, differently from the preliminary analysis (Neichel et al., 2014[\cite{neichel2014a}]), we basically never observed cases in which the discrepancy between observed and simulated data was not negligible. This tells us that a careful GeMS data reduction can provide an accurate estimate of the wind speed. Considering that the GeMS data reduction is manual, it is difficult to argue on potential causes of such an improvement with respect to Neichel et al. 2014. On the other side it is important the fact that such a good correlation indicates that the principle of GeMS measurement seems auto-consistent and reliable.

\begin{figure} [ht]
\begin{center}
\begin{tabular}{cc} 
\includegraphics[height=6cm]{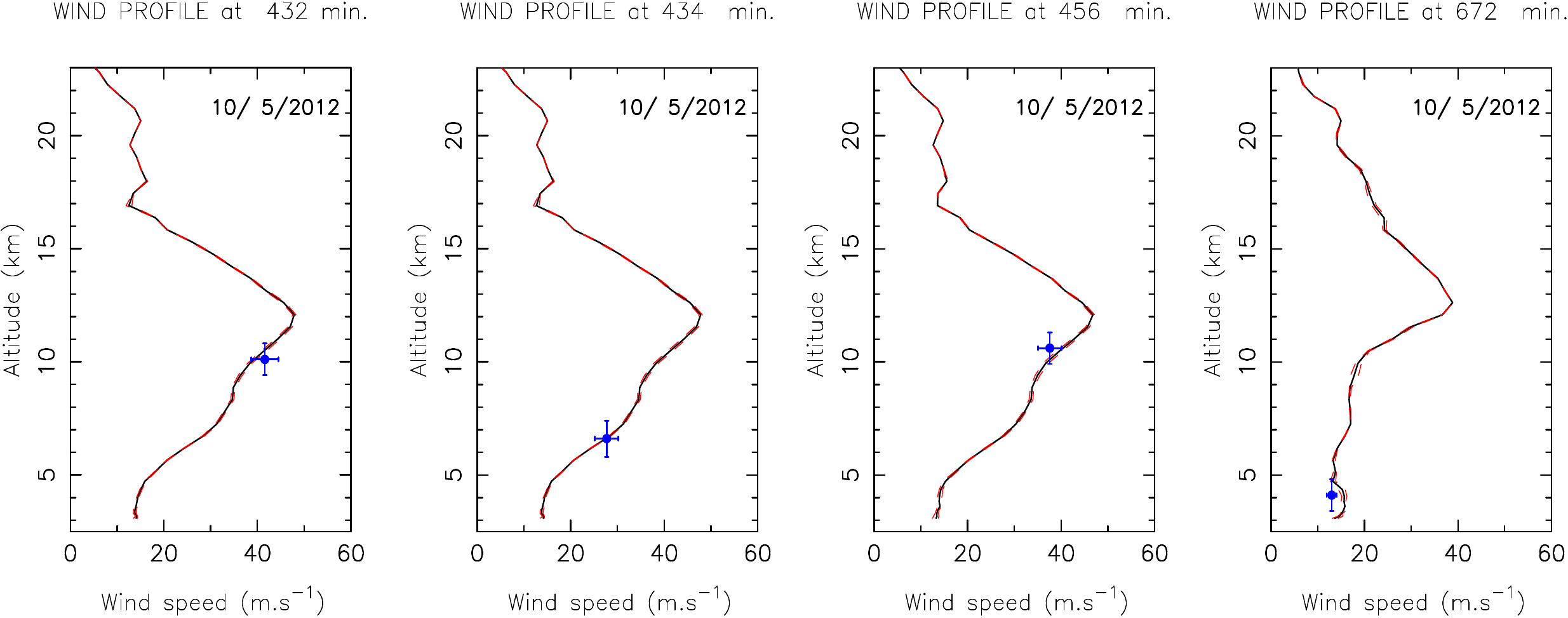}
\end{tabular}
\end{center}
\caption
{\label{fig:10052012} Example of wind speed estimates as provided by the model (vertical profile - black continuum line) and as estimated by GeMS (blue dots) during the same night. Red dashed lines represent the minimum maximum variation of the wind speed retrieved from the model in the $\pm$ 8 min range with respect to the closest instant in which the GeMS measurement took place. See text for errors bars (along x and z-axis). This night has been selected because, during the night, the AO system visibly  detected the wind speed at different heights always with a good level of correlation with the wind speed reconstructed by the model. }
\end{figure}

\begin{figure} [ht]
\begin{center}
\begin{tabular}{cc} 
\includegraphics[height=7cm]{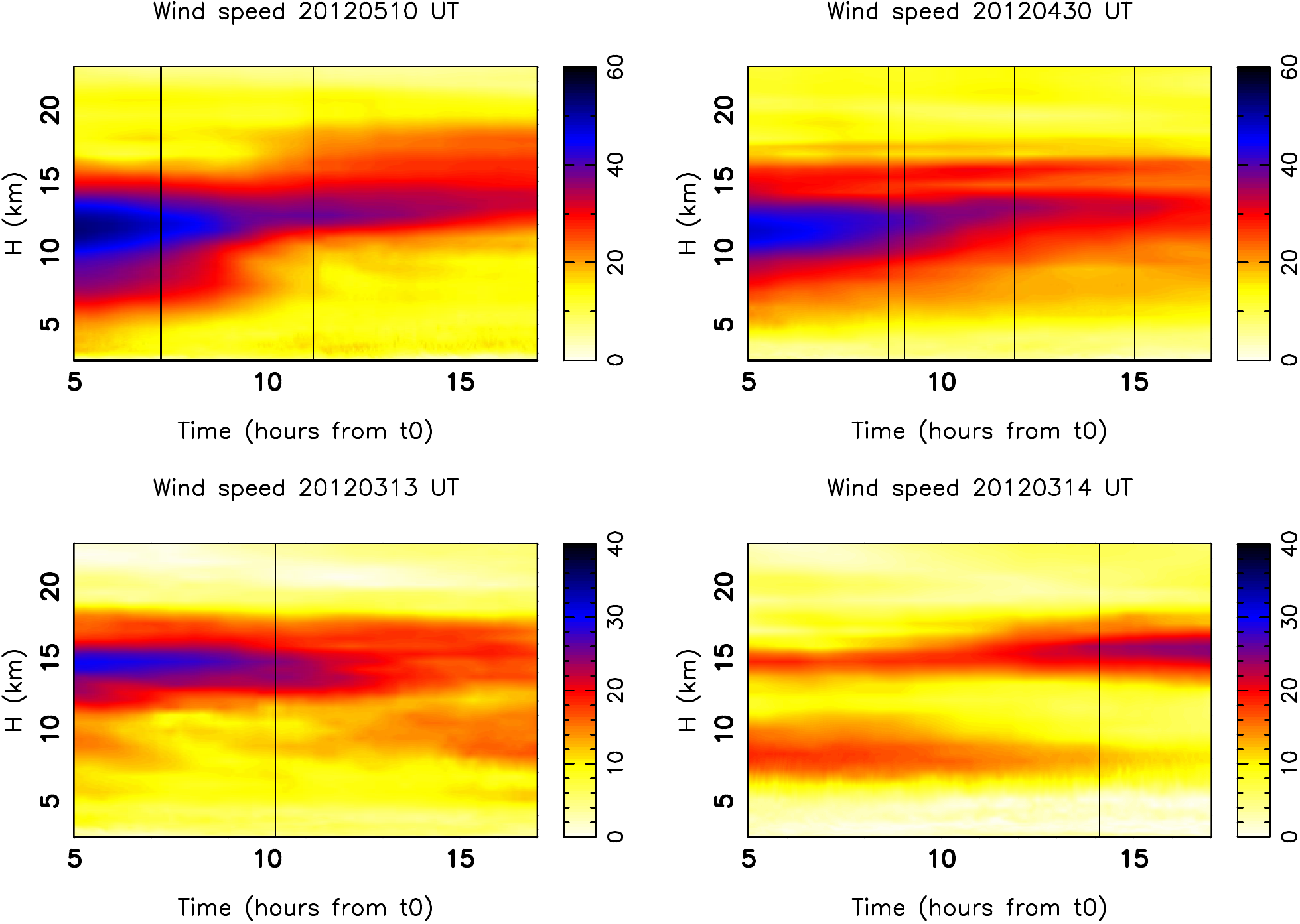}
\end{tabular}
\end{center}
\caption
{\label{fig:temp_evol1} Temporal evolution of the wind speed on the whole atmosphere as reconstructed by the Meso-Nh model all along a few nights. Black vertical lines are the instants in which measurements from GeMS are available.}
\end{figure}

\begin{figure} [ht]
\begin{center}
\begin{tabular}{cc} 
\includegraphics[height=7cm]{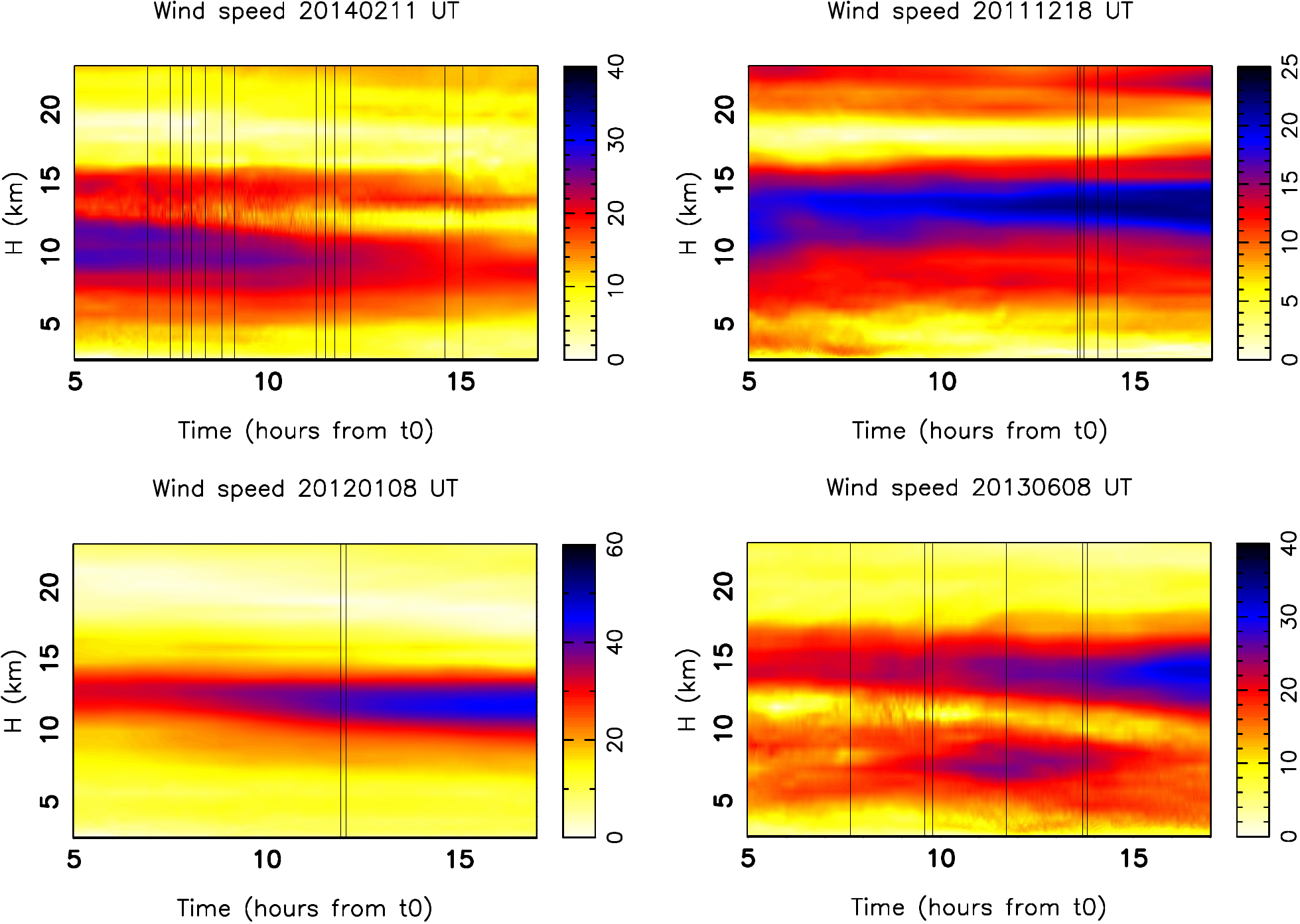}
\end{tabular}
\end{center}
\caption
{\label{fig:temp_evol2} 
As in Fig.\ref{fig:temp_evol1}}
\end{figure}

\begin{figure} [ht]
\begin{center}
\begin{tabular}{cc} 
\includegraphics[height=6cm]{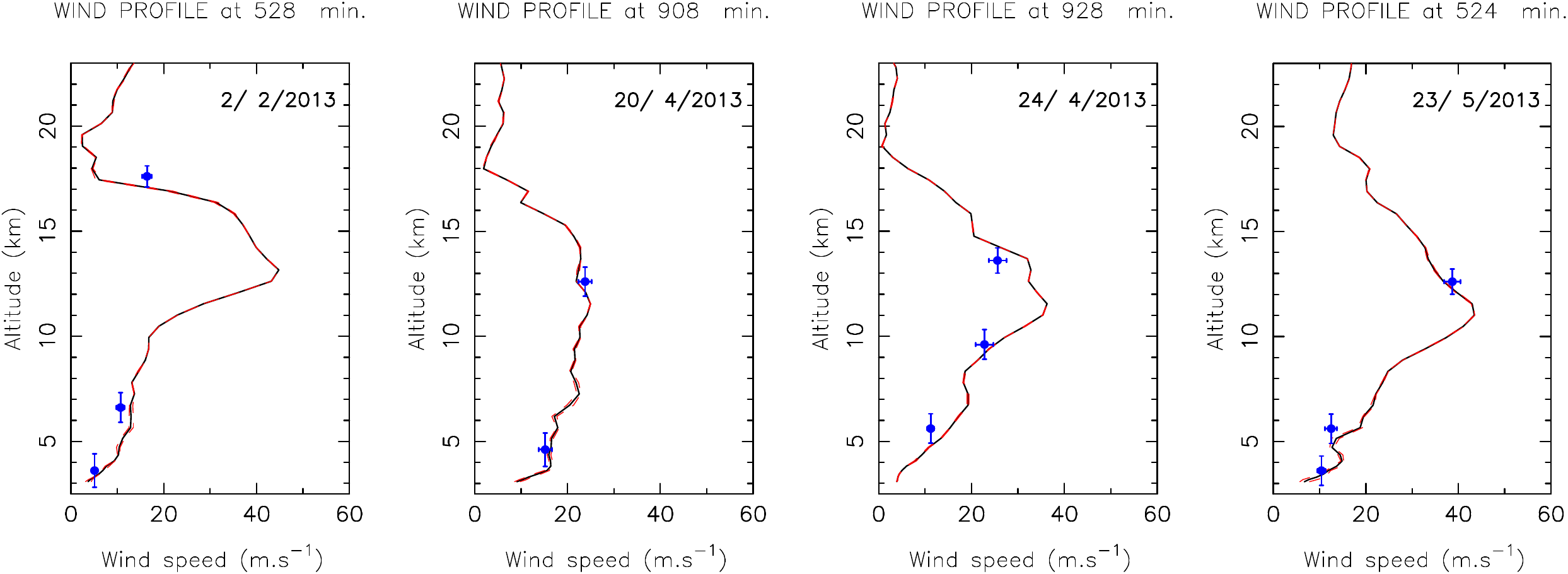}
\end{tabular}
\end{center}
\caption
{\label{fig:varie_wind_speed} 
Examples of wind speed estimated as provided by the model (vertical profile - black continuum line) and as estimated by GeMS (blue dots) during different nights. See Fig.\ref{fig:10052012} and text for errors bars (along x and z-axis).  We selected different nights in which GeMS detected the largest number of layers. Typically the system detects no more than three layers for the wind speed. }
\end{figure}

Figure \ref{fig:10052012} reports the example of one night in which GeMS detected wind speed associated to different layers moving at different heights (blue dots). GeMS estimates are visibly in perfect agreement with estimates provided by the model (black continuum lines). For what concerns the model outputs we selected the profiles closest to the instant in which observations took place with respect to the temporal sampling of 2 minutes of the model (black continuum line) and we displayed also the minimum and maximum values on a set of profiles taken at $\pm$ 8 minutes with respect to the black line i.e. $\pm$ 4 profiles (dashed red lines). 
In basically all cases the red dashed and black continuum lines are almost overlapped. This means that on a time scale of $\pm$ 8 minutes there are no major modifications of the wind as reconstructed by the model. However there are cases in which the wind speed spatial variability all along the 20~km and all along the individual nights is not negligible even if on longer time scales (see Fig.\ref{fig:temp_evol1} and Fig.\ref{fig:temp_evol2}). 
Figure \ref{fig:varie_wind_speed} shows the wind speed as measured by GeMS and as reconstructed by the model in other different nights in which GeMS detected the highest number of layers (three is the maximum number we detected in the whole sample). Also in these cases the agreement between GeMS and the model is very good. 
If we consider the whole sample of 43 night we can say that the error bar along z-axis of blue dots (GeMS data-set) is included in the [0.5, 0.8] km range and the error bar along x-axis is included in the [0.5, 3.8] ms$^{-1}$ range.                                                            

\begin{figure} [ht]
\begin{center}
\begin{tabular}{cc} 
\includegraphics[height=6cm]{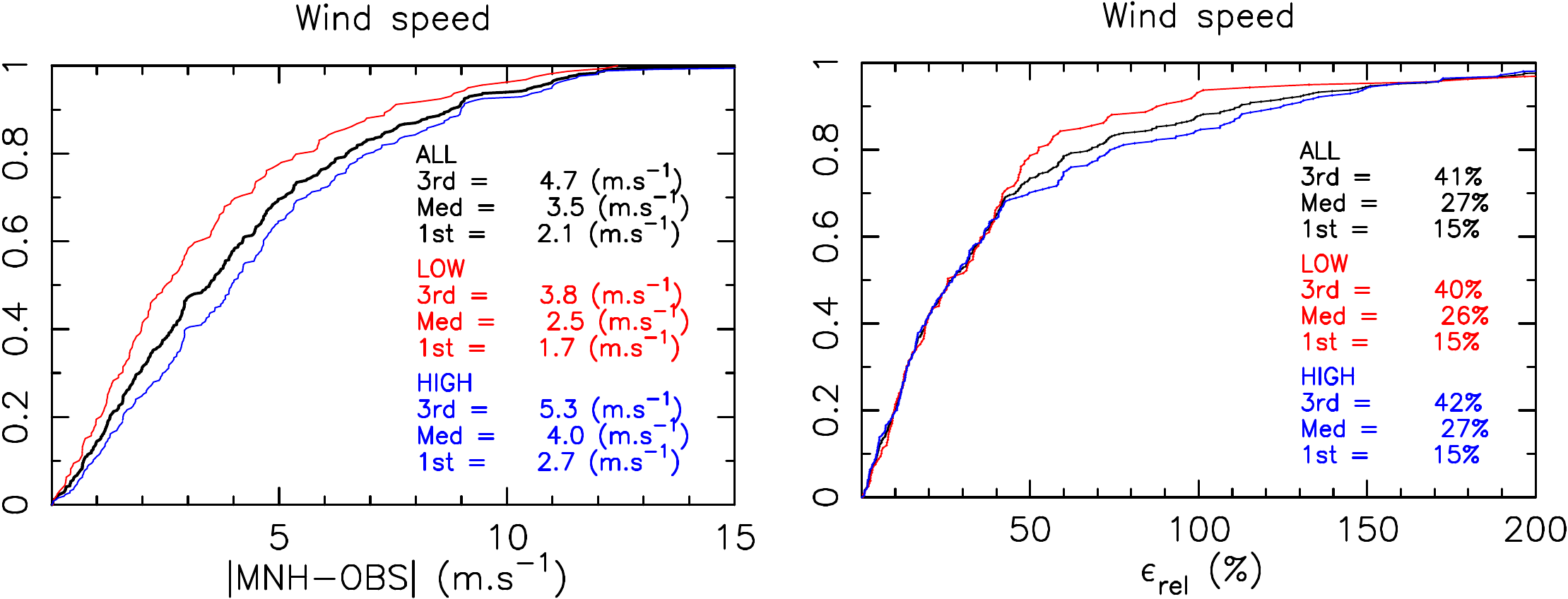}
\end{tabular}
\end{center}
\caption 
{\label{fig:cum_dist} 
Left: Cumulative distribution of the absolute difference of wind speed estimations provided by the model and by GeMS (sample of 400 points). Sample includes estimations related to the whole atmosphere [3~km, 18~km] a.s.l. (black line), values related to the low atmosphere i.e. [3~km, 5~km] range (red line), values related to the high atmosphere i.e. [5~km, 18~km] range (blue line). Right: Cumulative distribution of the relative error in the same vertical slabs. We have no GeMS estimates at heights h > 18~km.}
\end{figure} 

\begin{equation}
\left | V_{MNH}- V_{OBS}\right |
\label{eq1}
\end{equation}

\begin{equation}
\varepsilon_{rel} = \frac{\left | V_{MNH}-V_{OBS} \right |}{V_{MNH}} \times100\%
\label{eq2}
\end{equation}

Besides these particular cases of excellent performances, we can prove that the correlation between the GeMS measurements and model estimates of the wind speed is very satisfactory also in statistical terms. We compared indeed measurements with model outputs on the rich statistical sample of 43 nights corresponding to roughly 400 couples of values (measurement, model simulation).
Figure \ref{fig:cum_dist} shows the cumulative distribution of the absolute difference of the wind speed as estimated by GeMS and by the model (as Eq.\ref{eq1}) and of the relative error (as Eq.\ref{eq2}) in three different vertical slabs of the atmosphere: in the low part of the atmosphere i.e. in the [3~km, 5~km] range, in the high part of the atmosphere i.e. in the [5~km, 18~km] range and in the total atmosphere i.e. in the [3~km, 18~km] range. We have no GeMS detections of wind speed for heights higher than 18~km. All heights are calculated above the sea level (a.s.l.). The comparison has been performed with the following procedure: (1) we consider the estimate of GeMS $\pm$ the error bar along z-axis; (2) inside this vertical slab we calculate the average of the differences between the GeMS and the model estimates. Looking at Fig.\ref{fig:cum_dist} it is possible to conclude that the median value of the relative error in the low, high and total part of the atmosphere are respectively: 26\%, 27\% and 27\%. Similar results are obtained if we calculate the relative error with respect to the wind speed measured by GeMS instead than estimated by the model. If we look at Fig. \ref{fig:cum_dist} - left side, the median value of the difference | V$_{OBS}$ -V$_{MNH}$ | in the low, high and total atmosphere is respectively: 2.5 m$s^{-1}$, 4 m$s^{-1}$ and 3.5 m$s^{-1}$.

We highlight that in the first grid point, data from the model are considered starting from 500~m above the ground to eliminate the {\it grey zone} to be sure we are using the model in the same region of the atmosphere in which it has been validated. Between 30~m nd 500~m a comparison model vs. radio-soundings was meaningless due to orographic effects (see extended discussion in Masciadri et al. 2013[\cite{masciadri2013}]) and no conclusions could be retrieved. In the surface layer (roughly 30~m above the ground) the model showed to have a very good correlation with measurements. Validation has been done comparing model with measurements from sensors located at different heights (Lascaux et al. 2013, 2015[\cite{lascaux2013,lascaux2015}]) obtaining very good agreement. For this study we verified that there are no cases in which GeMS data falls below 500~m  i.e. below the grey zone. The comparison model-GeMS is therefore coherently done in the vertical slab of the first grid point for which the model was previously validated (Masciadri et al. 2013).

Table \ref{tab:res1} reports the bias, RMSE and $\sigma$ of the data-set in the whole atmosphere, in the low part of the atmosphere [3~km, 5~km] and in the high part of the atmosphere [5~km, 18~km]. $\sigma$ is the bias-corrected RMSE (see Eq.8 in Masciadri et al. 2013[\cite{masciadri2013}]). To facilitate the discussion of results, in the same table are reported also the correspondent values obtained comparing the model estimates with 50 radio-soundings[\cite{masciadri2013}]]. These data refer to the validation of the Meso-Nh model i.e. the study that lead us in considering Meso-Nh a reliable tool for the estimate of the wind speed and direction and the use of this model as a reference in this analysis. We can observe that bias, RMSE and $\sigma$ between GeMS and the Meso-Nh model are very satisfactory. They are just only slightly larger than the values obtained in Masciadri et al. (2013) study where model has been compared to radio-soundings. The difference is more evident for the $\sigma$ in the low part of the atmosphere (2.83~ms$^{-1}$ vs. 4.33~ms$^{-1}$). If we consider that GeMS has a much lower vertical resolution than the model we can consider these results very satisfactory. 

GeMS seems therefore to provide reliable estimates but, as we said, it is not so trivial the automation of such a system with an equivalent reliability. On the other side it should be much simpler the reading from the AO software of wind speed values provided by the model that has the advantages (1) to provide information of the wind speed on the whole range of 20~km, (2) with a temporal sampling of around 2 minutes and (3) to know that many hours in advance with respect to the observations.  
 
\begin{table}
\caption{Bias, RMSE and $\sigma$ between the GeMS and model estimates in two different part of the atmosphere calculated on the whole sample of 43 nights. Bias is calculated as |MNH-OBS|. $^{(*)}$Extracted from Masciadri et al. (2013)[\cite{masciadri2013}]}
\label{tab:res1}
\begin{center}
\begin{tabular}{l|ccc|ccc}
\hline
\multicolumn{1}{c}{} & \multicolumn{3}{c}{GeMS vs. Meso-Nh}&\multicolumn{3}{c}{Meso-Nh vs. Radio-soundings$^{(*)}$} \\
\multicolumn{1}{c}{} & \multicolumn{3}{c}{43 nights}&\multicolumn{3}{c}{50 nights} \\
\multicolumn{1}{c}{} & \multicolumn{3}{c}{(ms$^{-1}$)}&\multicolumn{3}{c}{(ms$^{-1}$)} \\
\hline
  & bias & RMSE & $\sigma$ & bias & RMSE &$\sigma$ \\
\hline
 LOW: [3,5]~km a.s.l.& 0.47 & 4.36 & 4.33 &  -1 & 3   &2.83 \\
 HIGH: [5,18]~km a.s.l. & -1.49 & 5.15 &  4.9& 0.6 & 3.5 & 3.45 \\
\hline
\end{tabular}
\end{center}
\end{table}

\begin{figure} [ht]
\begin{center}
\includegraphics[height=17cm]{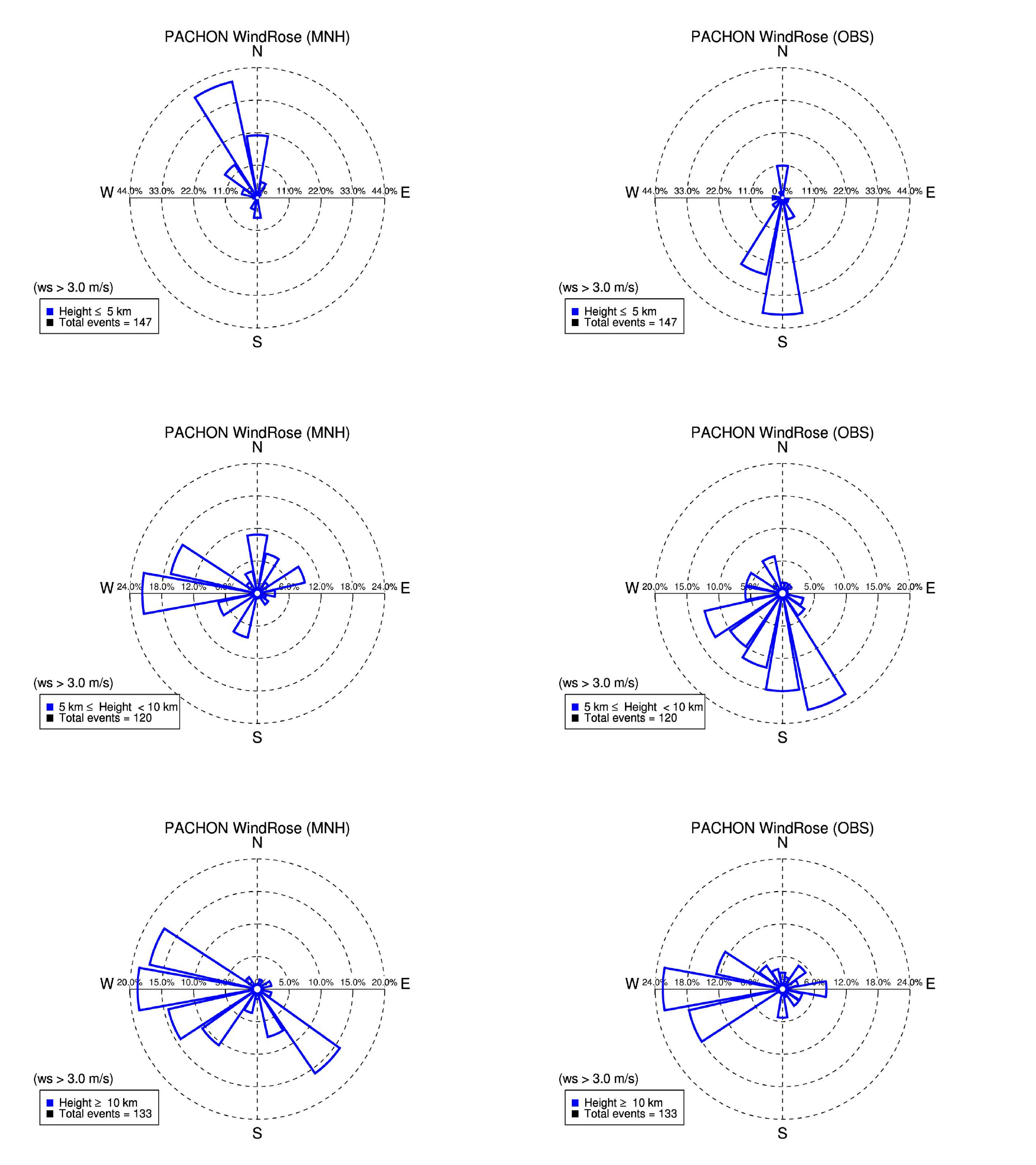}
\end{center}
\caption
{\label{fig:wind_rose_whole atm} Wind direction histogram calculated from the model (left) and GeMS (right) outputs in different vertical slabs (h < 5~km (top), 5~km < h < 10~km (centre), h > 10~km (bottom)) on the 43 nights sample (around 400 couples (measurement/model estimate)).}
\end{figure}

Besides the wind speed one might try to validate the GeMS ability in reconstructing the wind direction. Differently from the wind speed, if we compare the estimates provided by GeMS with those retrieved from the model we observed some not negligible uncertainties. Figure \ref{fig:wind_rose_whole atm} shows the wind rose as measured by GeMS (right column) and as provided by the model (left column). Histograms are divided in different vertical slabs: [3~km, 5~km] (top), [5~km, 10~km] (middle) and  [10~km, 18~km] (bottom). The largest discrepancies are observed in the low part of the atmosphere where the most frequent direction differs of almost 180$^{\circ}$. In the middle atmosphere the difference decreases to almost 90$^{\circ}$. In the high part of the atmosphere the agreement is better even if some important discrepancies are visible. It is not yet evident if the problem comes from the model or GeMS.

\begin{figure} [ht]
\begin{center}
\begin{tabular}{cc} 
\includegraphics[height=10cm]{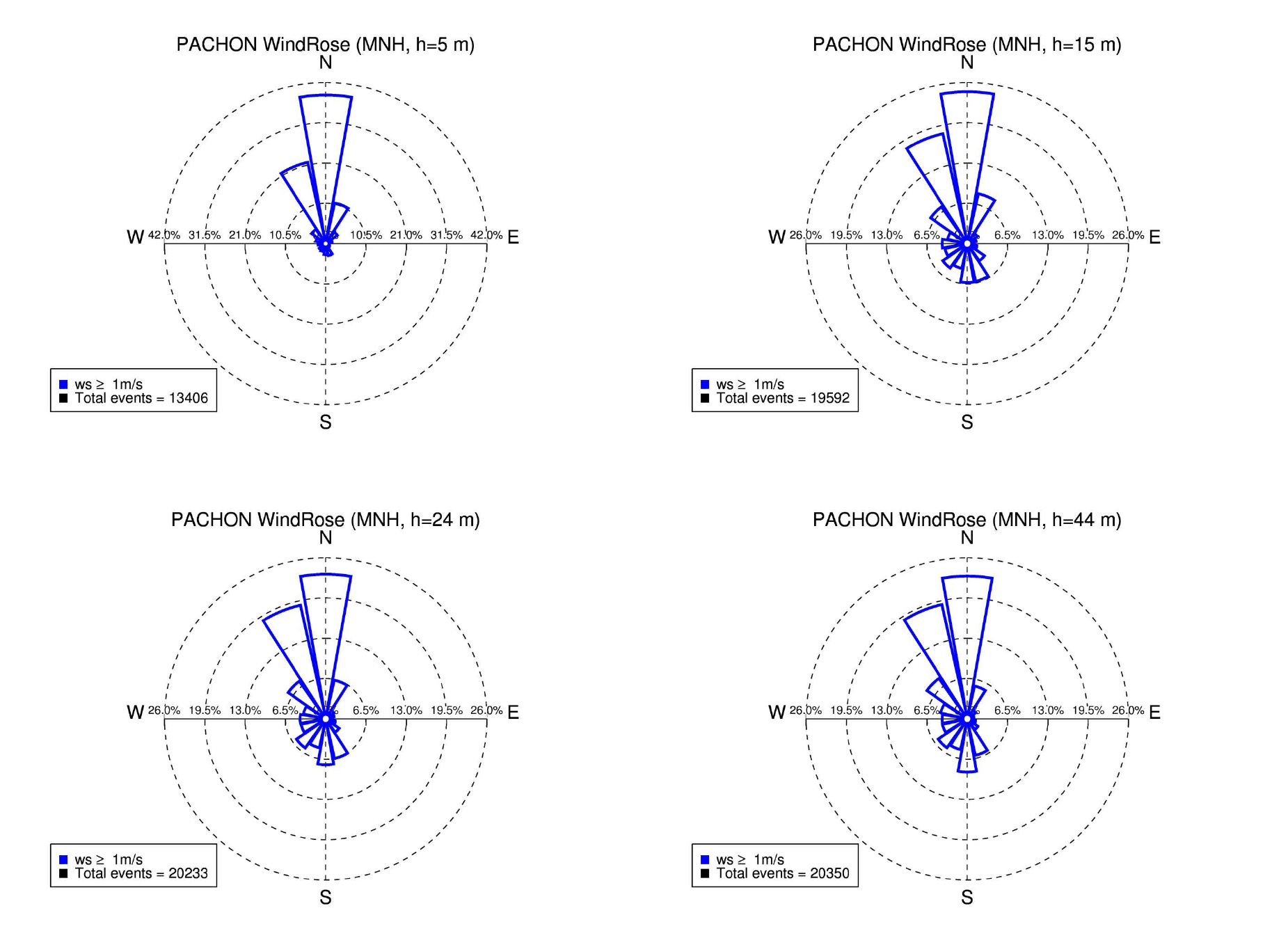}
\end{tabular}
\end{center}
\caption
{\label{fig:wind_rose_surf} Wind direction histogram as reconstructed by the Meso-NH model on the sample of 43 nights and related to the first four model levels above the ground (at 5~m, 15~m, 24~m and 44~m). We filtered out wind speed weaker than 1 ms$^{-1}$ because such a weak wind speed are in general highly dispersed and in any case not really useful for this application.}
\end{figure} 

The model has been previously validated above Cerro Paranal (Masciadri et al. 2013[\cite{masciadri2013}]) for the wind direction. In the range [5~km, 18~km] a.s.l. ([2.5~km, 15.5~km] a.g.l.) the medium RMSE between the model and the radio-soundings is very small (not larger than 10 degrees) and the model seems to provide a very good performances. It seems difficult that the model might have a completely wrong behaviour above Pachon in the middle-high atmosphere. In the surface layer (first 30 m) the model showed a good correlation with measurements (RMSE$_{relative}$ of the order of 19 \%) above Cerro Paranal. In the range of roughly 2 kilometers [3~km, 5~km] a.s.l. ([500~m, 2.5~km]), the so called 'grey zone', the RMSE can achieve some larger values (Masciadri et al. 2015[\cite{masciadri2015}]) even if well below 180$^{\circ}$. However there are no GeMS measurements in this vertical slab. We think therefore that this element is poorly influent in this discussion. Figure \ref{fig:wind_rose_surf} shows the wind rose of data as reconstructed by the model on the sample of 43 nights (400 estimates) above Pachon in the surface layer (roughly 45~m). Each panel report the most frequent directions from which the wind flows at the height correspondent to the first four model levels. We filtered out measurements associated to a wind speed weaker than 1~ms$^{-1}$ instead of 3~ms$^{-1}$ (as done in the whole atmosphere - see Fig.\ref{fig:wind_rose_whole atm}) because the wind speed close to the ground is weaker than in the rest of the atmosphere and we did not eliminate too many measurements from our sample. We can retrieve from Fig.\ref{fig:wind_rose_surf} that results are in perfect agreement with those obtained by Els et al., 2011[\cite{els2011}] performed in occasion of the characterisation of Cerro Pachon for the Large Synoptic Survey Telescope (LSST) showing a wind blowing mainly from the North-East and North-West. This test confirms the good model performances in the surface layer however, in any cases, all the 400 measurements from GeMS are associated to heights higher than 45~m. 

On the other side, knowing the principle on which the procedure to retrieve the wind speed for GeMS is based on, if the wind speed is well estimated by GeMS (and we proved that this is the case), it should be highly probably that the wind direction is well estimated too. We note that a set of trigonometric transformations have to be done to transform the wind direction expressed with respect to the CCD attached to the rotating telescope to the wind direction expressed with respect to the geographic references (wind blowing from the North corresponds to 0 degrees, wind blowing from East corresponds to 90 degrees). The software developed by one of us (AG) has been checked and no major problems are evident. We planned however to go deeper in the analysis of the wind direction to better discern the cause of this discrepancy.   


\section{CONCLUSIONS}
\label{sec:conclusions}

We presented a follow-up of a study started a couple of years ago[\cite{neichel2014a}] whose principle aim was to validate the principle used by GeMS to estimate wind speed and direction all along the 20~km of atmosphere above the ground by comparing GeMS estimates with those obtained with an atmospherical mesoscale model Meso-Nh previously validated. We compared wind speed and direction estimates retrieved from GeMS and Meso-Nh on a rich statistical sample of nights corresponding to roughly 400 estimates. GeMS data reduction and model simulations have been performed in a completely independent way. We proved that the wind speed estimates of GeMS are well correlated with the Meso-Nh estimates in the high as well as in the low part of the atmosphere with a medium RMSE of 5.15 ms$^{-1}$ and  4.36 ms$^{-1}$ respectively. The Meso-Nh model has a smaller RMSE with respect to radio-sounding (3.5 ms$^{-1}$ and 3 ms$^{-1}$) but, considering the lower vertical resolution of the GeMS, this result is not surprising. It is indeed a further element that indicates that the model is a good reference for the validation of new methods for the wind speed stratification. The correlation GeMS vs. model is not similarly good for the wind direction. Even if we observed some nice correlations in a few isolated cases, we observed important discrepancies whose cause is not, at present, yet evident. Problem might be due to some error in the sequence of geometric transformations necessary to translate the wind direction measurements done on sky and expressed with respect to the CCD attached to the rotating telescope in a wind direction expressed with respect to the geographic references (0$^{\circ}$ when the wind blows from the North, 90$^{\circ}$ when the wind blows from East) but further analyses are necessary to confirm or exclude this hypothesis. 
 
The conclusion is therefore that the principle used by GeMS to measure the wind speed is therefore consistent. Besides we observed that the GeMS data reduction is manual and its quality depends, visibly, on the careful with respect to which data reduction is performed. In any case it is hard to automatise this procedure. It is therefore appealing the solution to use, alternatively, the information provided by the Meso-Nh model by implementing it in the software of GeMS. Such an information is more complete in space and time and it can assure a systematic and automatic monitoring of the wind speed and direction. In the present study the vertical profiles had 62 levels (higher density close to the ground) and a 2 minutes temporal frequency all along the night.

We finally highlight that the fact that the principle of GeMS in retrieving the wind speed is strongly reliable has a positive effect on the principle of measurement of the \CN2 performed by GeMS. The method of wind speed is based on tracing the cross-correlation maps shifted from the origin where we have the auto-correlation map from which is retrieved the \CN2 profile.

\acknowledgments 
Based on observations obtained at the Gemini Observatory, which is operated by the Association of Universities for Research in Astronomy, Inc., under a cooperative agreement with the NSF on behalf of the Gemini partnership: the National Science Foundation (USA), the National Research Council (Canada), CONICYT (Chile), the Australian Research Council (Australia), Ministerio da Ciencia, Tecnologia e Inovaccao (Brazil) and Ministerio de Ciencia, Tecnologia e Innovacion Productiva (Argentina). Part of the numerical simulations have been run on the HPCF cluster of the European Center for Medium Range Weather Forecasts (ECMWF) using sources of the Project SPITFOT. Benoit Neichel acknowledges the French ANR program WASABI - ANR-13-PDOC- 0006-01. \\

\bibliographystyle{spiebib} 

\end{document}